\documentclass[pra,twocolumn,superscriptaddress]{revtex4}

\usepackage{graphicx}

\begin{document}

\title{Decoherence in a scalable adiabatic quantum computer}

\author{S. Ashhab}
\affiliation{Frontier Research System, The Institute of Physical
and Chemical Research (RIKEN), Wako-shi, Saitama, Japan}
\author{J. R. Johansson}
\affiliation{Frontier Research System, The Institute of Physical
and Chemical Research (RIKEN), Wako-shi, Saitama, Japan}
\author{Franco Nori}
\affiliation{Frontier Research System, The Institute of Physical
and Chemical Research (RIKEN), Wako-shi, Saitama, Japan}
\affiliation{Center for Theoretical Physics, CSCS, Department of
Physics, University of Michigan, Ann Arbor, Michigan, USA}

\date{\today}

\begin{abstract}
We consider the effects of decoherence on Landau-Zener crossings
encountered in a large-scale adiabatic-quantum-computing setup. We
analyze the dependence of the success probability, i.e. the
probability for the system to end up in its new ground state, on
the noise amplitude and correlation time. We determine the optimal
sweep rate that is required to maximize the success probability.
We then discuss the scaling of decoherence effects with increasing
system size. We find that those effects can be important for large
systems, even if they are small for each of the small building
blocks.
\end{abstract}

\maketitle

\section{Introduction}

The promise of enormous levels of speed up over classical
computing algorithms has stimulated research in the field of
quantum information processing, especially after the discovery of
a variety of concrete algorithms, including the factoring and
search algorithms \cite{Nielsen}. In the commonly studied
approach, to which we shall refer as sequential quantum computing
(SQC), the calculation is performed using a sequence of
pre-designed unitary operations on the quantum state of the
system. An alternative to SQC was proposed a few years ago, namely
adiabatic quantum computing (AQC) \cite{Farhi1,Farhi2}. The main
motivation for pursuing AQC is the idea that certain calculations
could be performed with speeds comparable to those obtainable with
SQC using a drastically different approach that avoids some of the
difficulties associated with SQC.

Calculations in AQC are performed as follows: one takes a given
quantum system and sets the external parameters such that the
system is guaranteed to relax to its ground state. One then slowly
varies those external parameters until the desired final set of
parameters is reached. The result of the calculation is then
encoded in the final quantum state, which should be the ground
state of the Hamiltonian at the end of the process. During this
adiabatic variation of parameters, a large number of avoided level
crossings are encountered, and the physics of Landau-Zener (LZ)
transitions applies \cite{Landau,OtherApplications}. The LZ
formula, which will be given below, states that if the time taken
to sweep across an avoided crossing is long compared to the
inverse of the gap in that crossing (we take $\hbar=1$), the
system remains in its ground state with a high degree of
certainty.

The fact that in AQC the system remains in its ground state
suggests, at least at first sight, that AQC is robust against
decoherence \cite{Childs,Roland}. In fact, that robustness is
generally thought of as being the single major advantage over SQC.
Recently it has been argued, however, that decoherence does set
limitations on AQC \cite{Roland,Shenvi,Sarandy}. In particular, if
the passage from the initial to the final state is done too
slowly, the success probability of the algorithm will be reduced
from the maximum obtainable value. In this paper we analyze the
{\it optimal} implementation of an AQC algorithm in the presence
of a noise source. We also discuss how decoherence effects
increase in importance with increasing system size. We show that
decoherence considerations can play a major role in determining
the optimal operation conditions of a scalable AQC setup.

This paper is organized as follows: In Sec. \ref{Sec:LZ} we
present the basic LZ problem. In Sec. \ref{Sec:GapScaling} we
briefly comment on the question of the scaling of the minimum gap
with system size. In Sec. \ref{Sec:LZDecoherence} we identify the
different regimes of robustness of AQC against decoherence, and we
analyze the optimal operation conditions for a prototypical AQC
algorithm in the presence of decoherence. In Sec.
\ref{Sec:DecoherenceScaling} we discuss the scaling of decoherence
effects with system size. Sec. \ref{Sec:Conclusion} presents some
concluding remarks.

\section{Landau-Zener problem without decoherence}
\label{Sec:LZ}

We start our discussion by introducing a prototypical example of
an AQC algorithm, namely the basic LZ problem. We therefore
consider a two state system, and we use the spin 1/2 language,
where the two states are called $\left|\uparrow\right\rangle$ and
$\left|\downarrow\right\rangle$. In the absence of coupling to the
environment, we take the time-dependent Hamiltonian:
\begin{equation}
\hat{H}(t) = - \frac{\Delta}{2} \hat{\sigma}_x - \frac{vt}{2}
\hat{\sigma}_z,
\end{equation}

\noindent where $\Delta/2$ is the tunnelling matrix element
between the states $\left|\uparrow\right\rangle$ and
$\left|\downarrow\right\rangle$, $v$ is the sweep rate of the
energy bias between the two states, and $\hat{\sigma}_{\alpha}$
are the Pauli spin matrices. The instantaneous (i.e., adiabatic)
two-level energy spectrum as a function of $vt$ is schematically
shown in the inset of Fig. 1. Note that the ground state and the
excited state at the degeneracy point (given by $vt=0$) are, with
the proper phase definitions, the symmetric and antisymmetric
superpositions of the eigenstates evaluated very far from the
degeneracy point. If the system is initially in its ground state
at $t\rightarrow -\infty$, the probability for the system to end
up in the new ground state at $t\rightarrow\infty$ is given by
\cite{Landau}:
\begin{equation}
P_{\rm LZ} = 1-\exp\left\{-\frac{\pi\Delta^2}{2v}\right\}.
\label{eq:LZ_formula}
\end{equation}

\noindent In particular, if the system crosses the degeneracy
region extremely slowly ($v\rightarrow 0$), the system is
guaranteed to end up in the new ground state. From now on, we
shall refer to the probability that the system ends up in the new
ground state as the success probability, since that situation
represents a successful run of this prototypical AQC algorithm.

\section{Scaling of minimum gap with system size}
\label{Sec:GapScaling}

Before going into any details regarding decoherence, it is worth
mentioning here one of the most relevant open questions in the
study of AQC, namely the dependence of the minimum gap between the
ground state and first-excited state on the system size
\cite{GapDependence}. Since the size of that gap sets an upper
bound on the allowed sweep rate, an increasingly small gap could
deem an AQC algorithm ineffective to solve a given problem,
especially in the case of an exponentially decreasing gap.
Although that scenario would also make the algorithm more
susceptible to decoherence, the scaling of the gap is not directly
related to the present discussion. We shall therefore not dwell
upon that question in this paper, and we shall leave any
dependence of the minimum gap on system size implicit.
Incorporating a given dependence into our results can be done
straightforwardly.

\section{Landau-Zener problem with decoherence}
\label{Sec:LZDecoherence}

Let us start by presenting an argument that is sometimes used to
suggest robustness of AQC against decoherence. We divide the noise
effects into high-frequency and low-frequency contributions.
High-frequency noise is responsible for relaxation processes
(i.e., transitions between different energy levels), whereas
low-frequency noise is responsible for dephasing processes. If we
assume that the temperature is lower than the minimum gap
encountered while running the algorithm \cite{GapAssumption}, the
excitation rate will always be small in comparison to the
de-excitation rate, and the system will relax to the new ground
state at the end of every LZ crossing if necessary. High-frequency
noise can therefore be neglected. Now, since the system is always
in an eigenstate of the Hamiltonian, namely the ground state,
dephasing is irrelevant. Low-frequency noise, which describes
dephasing processes, can therefore be neglected as well. One would
therefore conclude that AQC is robust against decoherence.

Given that the above argument gives strong support to AQC over
SQC, we now discuss in some detail its applicability in different
possible situations. An important point to note here is that the
argument implicitly uses perturbation-theory results regarding
relaxation and dephasing processes. That approach is valid only
when the noise amplitude is small compared to the qubit energy
scales. In particular, if the assumption of small amplitudes in
the noise signal is abandoned, the argument breaks down. As we
shall discuss in Sec. \ref{Sec:DecoherenceScaling}, this breakdown
seems to be the case for a scalable AQC system. Furthermore,
relaxation between macroscopically distinct quantum states after
the LZ crossing should be negligible.

A number of different approaches have been used to study the
effects of decoherence on the LZ transition probability
\cite{Sarandy,Kayanuma,Ao,Shimshoni,Nishino}. Although those
approaches are based on different underlying assumptions, they all
produce similar qualitative results (note that they have different
predictions regarding certain details). In particular, all of them
predict the possibility of having a maximum in the success
probability as a function of sweep rate (see Fig. 1).

Since we shall treat a number of qualitatively different cases, it
would be difficult to use a single model to describe the effects
of the environment on the success probability. We shall therefore
use two different models: one with a classical noise signal and
one with an environment of quantum modes. In addition, we shall
use thermodynamics principles when necessary.
\begin{figure}[ht]
\includegraphics[width=8.5cm]{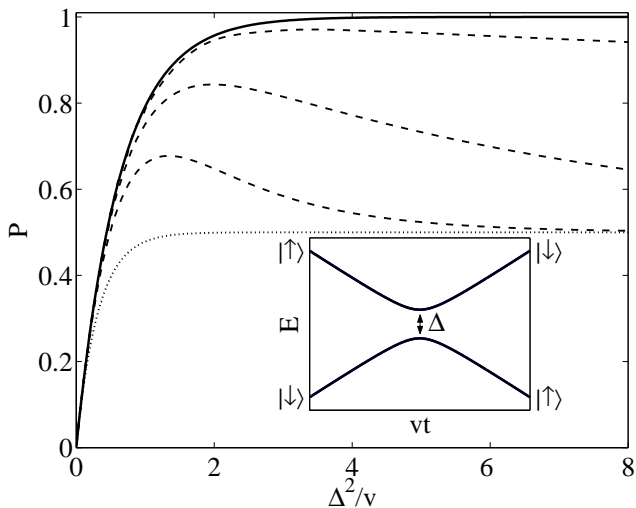}
\caption{\label{fig1} Success probability $P$, i.e. the
probability to end up in the new ground state after a Landau-Zener
crossing, as a function of $\Delta^2/v$, where $\Delta$ is twice
the tunnelling matrix element, and $v$ is the energy-bias sweep
rate. The solid line corresponds to the case of no decoherence.
The dashed lines correspond to the case of intermediate levels of
decoherence (essentially using the classical-noise model); the
curves were obtained following Ref. \cite{Kayanuma} with dephasing
rate $\Gamma_2(t\rightarrow\pm\infty) = \Delta/200$, $\Delta/20$,
and $\Delta/5$. The dotted line corresponds to the limit of
infinitely strong decoherence. Inset: Schematic view of the
instantaneous two-level energy spectrum as a function of the
energy bias $vt$.}
\end{figure}

Before analyzing the effects of the environment on the system
dynamics, we must specify the system operator involved in the
system-environment coupling. In the simple two-level problem that
we are considering, that operator must be one, or a combination,
of the Pauli matrices, assuming the coupling is described by a
product of a system operator and an environment operator. We note
that away from the crossing region coupling through the operator
$\hat{\sigma}_z$ only causes dephasing, whereas coupling through
the operators $\hat{\sigma}_x$ and $\hat{\sigma}_y$ causes
relaxation. In a macroscopic system, relaxation processes between
macroscopic states are generally exponentially small. We therefore
approach the problem at hand by taking the system introduced in
Sec. \ref{Sec:LZ} and adding a decoherence term that couples to
the system through the operator $\hat{\sigma}_z$. Although in
general more complex models (i.e., many-level models) must be used
to obtain a more detailed description of the effects of noise on a
large AQC system, the arguments given below provide an initial
understanding of some of the main mechanisms involved in the
problem.

\subsection{Classifying the noise according to amplitude and correlation
time}

We divide our discussion into four cases, determined by the
following procedure: we take a noise signal with characteristic
amplitude $A$ (in energy units) and correlation time $\tau$. We
note that the power spectrum of the noise signal would be
characterized by a (maximum) frequency $\omega_{\rm max}$ that is
related to the correlation time $\tau$ by $\omega_{\rm max}\equiv
1/\tau$. The noise power spectrum is then of order
$A^2/\omega_{\rm max}$ up to frequency $\omega_{\rm max}$ and
decreases to zero at higher frequencies. Note also that if the
noise signal has a non-zero average value, we define $A$ as the
deviation from that average value. Depending on whether $A$ is
smaller or larger than the gap $\Delta$, the noise is
characterized as low- or high-amplitude noise. Similarly,
depending on the relation between $\omega_{\rm max}$ and $\Delta$,
the noise is characterized as having short or long correlation
time.

\subsubsection{Low-amplitude noise with short correlation time}
\label{Subsubsec:HFLA_noise}

We start with this case because it allows the use of the simple
perturbation-theory results mentioned above. We focus on
relaxation processes, because pure dephasing processes cannot have
a larger effect than relaxation processes (note that relaxation
dynamics automatically contains dephasing), and therefore
including those cannot change the qualitative results we shall
give below. We also neglect de-excitation processes for a moment.
Away from the degeneracy region, the transition rate from the
ground state to the excited state is negligible because the noise
couples to the system through the operator $\hat{\sigma}_z$, which
is almost parallel to the system Hamiltonian. We therefore focus
on the dynamics when the system is close to the degeneracy point.
Since the noise power spectrum extends to frequencies higher than
$\Delta$, one finds the excitation rate from the ground state to
the excited state around the degeneracy point to be
\begin{equation}
\Gamma_{0\rightarrow 1} \sim \frac{A^2}{\omega_{\rm max}},
\end{equation}

\noindent which is essentially the noise power spectrum at the
transition frequency. One therefore straightforwardly finds that
the time spent traversing the LZ crossing must be shorter than
$1/\Gamma_{0\rightarrow 1}\sim \omega_{\rm max}/A^2$ if the noise
effects are to be minimized. Combined with the condition that the
traversal time must be larger than $1/\Delta$, one can determine
the ideal range of sweep rates for optimal AQC operation. If we
take the noise-induced excitation probability to be
\begin{equation}
P_{\rm excited \ by \ noise} \sim \frac{A^2\Delta}{\omega_{\rm
max} \; v}
\end{equation}

\noindent and the LZ transition probability to be
\begin{equation}
P_{\rm excited \ by \ LZ} \sim
\exp\left\{-\frac{\pi\Delta^2}{2v}\right\},
\end{equation}

\noindent and we minimize the sum of those two terms, we find that
the optimal value of $v$ is roughly given by
\begin{equation}
v_{\rm optimal} \sim \frac{\Delta^2}{\ln(\omega_{\rm max} \;
\Delta/A^2)}.
\end{equation}

\noindent Similarly, one can estimate that the maximum achievable
success probability will be $1-P_{\rm failure}$, with
\begin{equation}
P_{\rm failure} \sim \frac{A^2}{\omega_{\rm max} \; \Delta}.
\label{eq:P_failure}
\end{equation}

\noindent Note that the optimal sweep rate $v_{\rm optimal}$
depends logarithmically on the noise amplitude. That result
implies that $v_{\rm optimal}$ can be only a few times smaller
than $\Delta^2$ even if the noise power spectrum is orders of
magnitude smaller than $\Delta$.

We note here that if one is considering the case where the
temperature $k_BT$ is smaller than the gap $\Delta$, the
excitation rate will be smaller than the de-excitation rate by a
factor $\exp\{-\Delta/k_BT\}$, and the thermal-equilibrium
occupation probability of the excited state is given by
$1/(1+\exp\{\Delta/k_BT\})$. Therefore the above results apply
only if the expression in Eq. (\ref{eq:P_failure}) is smaller than
the thermal-equilibrium occupation probability. Otherwise, one
must take the de-excitation rate into account. One then finds that
the maximum obtainable success probability is given by the
thermal-equilibrium value $1/(1+\exp\{-\Delta/k_BT\})$, and it is
achieved using a slow sweep such that thermal equilibrium is
reached.

\subsubsection{Low-amplitude noise with long correlation time}
\label{Subsubsec:LFLA_noise}

Since the noise amplitude is small, one can still think of the
noise effects in terms of the transition rate from the ground
state to the excited state. The transition rate in this case can
be thought of as a high-order process \cite{Usuki}. Thinking of
the noise as a harmonic-oscillator bath, we find that an
$n$th-order process is required to excite the two-level system,
with $n={\rm Int}(\Delta/\omega_{\rm max})+1$, and the function
${\rm Int}(x)$ gives the highest integer smaller than $x$. For a
more concrete visualization, one can think of a photon bath, such
that the sum of $n$ photon energies is required to excite the
system from the ground state to the excited state. The transition
rate is therefore
\begin{equation}
\Gamma_{0\rightarrow 1} \sim \frac{A^2}{\omega_{\rm max}}\times
\left(\frac{A}{\Delta}\right)^{2n-1}.
\end{equation}

\noindent The above expression for the excitation rate suggests
that for the noise-driven excitation probability to be negligible
the time taken to traverse the LZ crossing must be smaller than
$1/\Gamma_{0\rightarrow 1} \sim (\omega_{\rm max}/A^2) \times
(A/\Delta)^{1-2n}$. Given that $A$ is smaller than $\Delta$, the
upper bound on crossing time above is much larger than $1/\Delta$.
This case is therefore the ideal case for performing AQC, allowing
a high success probability when an appropriately chosen sweep rate
is used. An estimate of the optimal sweep rate and the maximum
achievable success probability can be obtained similarly to what
was done in Sec. \ref{Subsubsec:HFLA_noise}. In this case one
finds
\begin{eqnarray}
v_{\rm optimal} \sim \frac{\Delta^2}{\ln(\omega_{\rm max} \;
\Delta^{2n}/A^{2n+1})}
\\
P_{\rm failure} \sim \frac{A^2}{\omega_{\rm max} \; \Delta} \times
\left(\frac{A}{\Delta}\right)^{2n-1}.
\end{eqnarray}

Note that the characteristic noise frequency $\omega_{\rm max}$
cannot be larger than the temperature $k_BT$, so that the lowest
possible value of $n$ is roughly
\begin{equation}
n_{\rm min} \sim {\rm Int}\left(\frac{\Delta}{k_BT}\right).
\end{equation}

\noindent Note also that if the expression for $P_{\rm failure}$
above is larger than $1/(1+{\rm exp}\{\Delta/k_BT\})$ the optimal
approach would be a slow sweep such that thermal equilibrium is
reached during the crossing.

\subsubsection{High-amplitude noise with long correlation time}
\label{Subsubsec:LFHA_noise}

We now take a slowly varying classical noise signal with an
amplitude larger than $\Delta$ (note that the slowness is
determined by comparison to the inverse of the gap). We also take
the system to be biased close to or at the degeneracy point. Since
the amplitude of the noise signal is larger than the gap, one
cannot use perturbation-theory results to describe transitions
between the different eigenstates. Instead, one can now think of
the noise signal as repeatedly driving LZ crossings, with
noise-driven sweep rate
\begin{equation}
v_{\rm env} \sim \frac{A}{\tau} \sim A \; \omega_{\rm max}.
\end{equation}

\noindent The LZ transition probability [$1-P_{\rm LZ}$, with
$P_{\rm LZ}$ given by Eq. (\ref{eq:LZ_formula})] with sweep rate
$v_{\rm env}$ is therefore not necessarily small, even if
$\omega_{\rm max}$ is much smaller than the gap. In particular,
the transition probability in an environment-induced LZ crossing
is (very roughly) given by
\begin{equation}
P_{\rm excited \ by \ env.-ind. \ LZ} \ \sim \
\exp\left\{-\frac{\pi\Delta^2\tau}{2A}\right\}.
\end{equation}

\noindent Given enough time, the system will therefore reach a
state where both eigenstates have equal occupation probabilities.
However, because of the exponential dependence of the transition
probability on the noise parameter, one can say that if the
condition $\pi\Delta^2\tau/2A \gg 1$ is satisfied, the
environment-induced LZ transition probability will be small enough
that a high success rate is always achievable with a properly
chosen value of $v$. The above criterion therefore provides the
condition for high-amplitude noise to have a negligible effect on
the success probability.

One might now raise the following possibility: taking a LZ
situation where the parameters are swept across the degeneracy
region, one can estimate that the number of noise-driven crossings
is of the order of $A/v\tau$. Therefore, if the sweep rate $v$ is
substantially larger than $A/\tau$, no environment-driven
crossings will occur, suggesting that it might be possible to
avoid environment-driven LZ transitions even if the condition
$\pi\Delta^2\tau/2A \gg 1$ is not satisfied. It is straightforward
to verify, however, that in order to do so one would require a
value of $v$ larger than $\Delta^2$. That situation would result
in a high bias-driven LZ transition probability and, therefore, a
low success probability.

\subsubsection{High-amplitude noise with short correlation time}
\label{Subsubsec:HFHA_noise}

In this case one can follow the above arguments for the
high-amplitude, low-frequency noise. Using the expressions of Sec.
\ref{Subsubsec:LFHA_noise}, one immediately finds the intuitively
obvious result that the success probability is 50\% for low sweep
rates and is smaller than that value for fast sweep rates (see the
dotted line in Fig. 1). Note that the value 50\% describes the
case where the two eigenstates have equal occupation probabilities
at the end of the process. Note also that since we have in mind
macroscopic states, we neglect the possibility that the system
could relax to the ground state long after the LZ crossing.

\section{Scalable system}
\label{Sec:DecoherenceScaling}

We now turn to the question of how decoherence effects scale with
system size in an AQC setting with a large number $N$ of qubits
(we use the typical picture of two-state qubits).

We have discussed in Sec. \ref{Sec:LZDecoherence} that for
large-amplitude noise one must think of different decoherence
mechanisms than the usual perturbation-theory relaxation and
dephasing mechanisms. We therefore consider the question of how
the noise amplitude scales with system size \cite{GapAssumption}.
In relation to that discussion, it is useful to classify LZ
crossings according to the number of qubits that change their
state during the transition. That criterion is related to, but
clearly distinct from, the question of quantifying how macroscopic
a quantum state is. There has not been any unambiguous and
universally accepted formulation of such a quantity. Following
Ref. \cite{Leggett} we use a common-sense definition rather than
trying to formulate an operational one, which seems to be a
formidable task. The definition is then relatively simple: a given
LZ crossing can be referred to as an $M$-qubit crossing if $M$
qubits change their state with the other qubits in the system
experiencing negligible changes. We can then speak of few-qubit
and many-qubit crossings. The former refers to LZ crossings of the
$N$-qubit system where only a few (say, up to four) qubits change
their state, even if the total number of qubits in the system is
macroscopic. The other type of LZ crossings that can occur during
the operation of an algorithm are many-qubit crossings. In those
crossings the number of qubits that change their state is of order
$N$.

In order to demonstrate the above-mentioned distinction between
classifying quantum states and classifying LZ crossings, take the
plausible scenario of AQC where one starts with a quantum state
that contains negligible multi-qubit entanglement and reaches a
quantum superposition of macroscopically distinct states during
the calculation. Although the quantum state becomes a macroscopic
one, it is not necessarily the case that any many-qubit crossings
must have been encountered (think for example of a macroscopic
quantum state generated by repeatedly performing two-qubit CNOT
gates). One should also note that even if the system is in a
superposition of macroscopically distinct states, it can still
undergo few-qubit LZ transitions. Those transitions would most
likely occur in one or some of the branches corresponding to the
different macroscopic states.

We now take an $M$-qubit LZ crossing. The size of the degeneracy
region is of the order of the gap $\Delta$. In a system with a
large number of degrees of freedom, one can still say that the
crossing region is defined by being within distance (in units of
bias parameters) $\Delta$ in the relevant $M$ directions from the
degeneracy point, i.e. the point where the gap takes its smallest
value along the path of the AQC algorithm.

If the noise signal on a single qubit moves the system away from
the bias point by a distance of order $\delta$, the sum of the
noise signals acting on the $M$ qubits moves the system away from
the bias point by a distance of order $\sqrt{M}\delta$. We now
take a system at or near the degeneracy point. If the total
deviation caused by the noise is smaller than the width of the
crossing region, which is of the order of $\Delta$, we can use the
arguments of Sec. \ref{Sec:LZDecoherence} to say that low
frequency noise can be neglected in the sense that it cannot
excite the system from its ground state.

In the opposite case, i.e. when the amplitude of the total noise
signal is larger than $\Delta$, one has to worry about
environment-driven LZ transitions. Using the results of Sec.
\ref{Sec:LZDecoherence}, we find that a rough estimate of the
probability for the system to be excited from its ground state
during a single typical (environment-driven) crossing is given by
\begin{equation}
P_{\rm excited \ by \ env.-ind. \ LZ}
\sim\exp\left\{-\frac{\pi\Delta^2\tau}{2\sqrt{M}\delta}\right\}.
\end{equation}

\noindent Note that the exponential dependence of the excitation
probability on the noise signal means that the above expression
should be thought of as an optimistic estimate; the true
excitation probability will probably be higher, depending on the
temporal behaviour of the noise signal. Using the results of Sec.
\ref{Sec:LZDecoherence}, the criterion on the tolerable
single-qubit noise can now be given by
\begin{equation}
\delta \ll \frac{\Delta^2\tau}{\sqrt{M}}.
\end{equation}

The probability that the noise signal will excite the system out
of its ground state therefore depends on the typical value of $M$
characterizing the LZ crossings that are encountered during the
algorithm. Given the scaling of the excitation probability with
$M$, it is highly desirable to follow a path in the
many-dimensional parameter space such that many-qubit LZ crossings
are avoided. This principle can therefore remain as a major
consideration in designing AQC algorithms, even if the minimum-gap
problem discussed in Sec. \ref{Sec:GapScaling} is solved.

It is not clear whether in a general AQC problem a path that
avoids all many-qubit LZ crossings exists. The 3-SAT problem,
which is a commonly studied potential application of AQC
\cite{Hogg}, provides an example where it seems impossible to find
such a path. In that problem one looks for a classical state of
the qubits such that a large number of 3-qubit logical conditions
are satisfied, e.g. the Boolean condition ``(qubit 5 and qubit 24)
or qubit 57". In the plausible scenario where one configuration
satisfies all the logical conditions but a large number of other,
macroscopically distinct configurations violate only a few
conditions, a quantum superposition involving a large number of
macroscopically distinct configurations must be retained until
near the end of the calculation, as they are eliminated slowly
with the testing of more and more conditions. At that point it
would require a many-qubit LZ crossing to eliminate those last
surviving near-solutions in favor of the unique solution of the
problem. The 3-SAT problem therefore appears to be one where
decoherence can be a major obstacle. The fact that the path of an
AQC algorithm is designed without knowing the quantum state that
will exist at each point in the algorithm raises similar doubts
about the possibility of {\it a priori} guessing the best path to
follow in a general problem.

The above arguments therefore raise questions that must be
answered in designing an AQC approach in a macroscopic setup.
Until those questions are answered, it is not clear to what extent
AQC is less susceptible to noise than SQC, especially given the
condition that we found above requiring the noise signal to
decrease with increasing system size \cite{GapAssumption}.

Even if achieving the ground state is not possible, e.g. because
of decoherence or a small minimum gap, a new promising proposal
notes that finding a near-solution can, under certain conditions,
be considered a success of the algorithm \cite{Zagoskin}. Because
a high success probability (in the sense of Sec. \ref{Sec:LZ}) is
not required, that approach could be more robust against
decoherence.

It is also worth noting here that we have used the simple model of
a two-state LZ problem, which represents a prototypical AQC
algorithm. The number of degrees of freedom in a large AQC setup
increases with system size. More complex models will be required
in order to both analyze the effects of noise and determine the
optimal path in those many-dimensional problems. Reaching a better
understanding of the structure of the energy manifolds in these
many-dimensional systems is therefore highly desirable.

\section{Conclusion}
\label{Sec:Conclusion}

We have analyzed the effects of noise on a prototypical AQC
algorithm, namely the LZ problem. We have found general principles
that determine the robustness of the algorithm against noise
sources with a variety of properties according to their amplitude
and correlation times. We have also determined the ideal operation
conditions that are required to maximize the success probability,
and we have analyzed the scaling of noise effects with system
size. Our results provide guidelines for the optimal
implementation of an AQC algorithm and raise questions that must
be answered before determining the suitability of AQC to tackle a
given problem. Given the promise of AQC as an alternative approach
to achieve extremely high-speed computation, we believe that our
results will contribute to a better understanding of that
approach, towards which initial experimental steps have already
been taken \cite{Steffen,Izmalkov}.

\begin{acknowledgements} This work was supported in part by the
National Security Agency (NSA), the Laboratory for Physical
Sciences (LPS) and the Army Research Office (ARO); and also by the
National Science Foundation (NSF) grant No.~EIA-0130383. One of us
(S. A.) was supported by the Japan Society for the Promotion of
Science (JSPS).
\end{acknowledgements}


\begin{thebibliography}{99}
\bibitem{GapAssumption} Note that we are leaving the gap dependence
on system size implicit. We therefore do not deal with any
difficulties associated with a shrinking minimum gap.
\end{thebibliography}
\end{document}